\documentclass[showpacs,twocolumn,preprintnumbers,amsmath,amssymb] {revtex4}
\usepackage{graphicx,color,epsfig}
\usepackage{dcolumn}
\usepackage{bm}

\begin{document}
\topmargin 0.2cm

\title{ \bf  Quark Gluon Plasma (QGP) evolution under loop corrections }

\author{K K Gupta$^1$, Agam K Jha$^2$, S. Somorendro Singh$^3$\footnote{email:sssingh@physics.du.ac.in} }
\affiliation {$^1$Department of Physics, Ramjas College, University of Delhi, Delhi - 110007, India \\
$^{2}$ Department of Physics, Kiroli Mal College, University of Delhi, Delhi - 110007, India \\
$^{3}$ Department of Physics and Astro-Physics, University of Delhi, Delhi - 110007, India \\}        
\begin{abstract}
 We review free energy evolution of QGP (Quark-gluon plasma) under
zero-loop, one loop and two loop corrections in the mean field potential. 
The free energies of QGP under the comparison of zero-loop and loop corrections 
of the interacting potential among the quarks, anti-quarks and gluons are
shown. We observe that the formation of stable QGP droplet is dependent 
on the loop corrections with the different parametrization values of fluid.
With the increase in the parametrization value, stability of droplet 
formation increases with smaller size of droplet. This indicates that 
the formation of QGP droplet can be signified more importantly by 
the parametrization value like the Reynold number in fluid dynamics. 
It means that there may be different phenomenological parameter to 
define the stable QGP droplet when QGP fluid is studied under 
loop corrections.  

\end{abstract}
\vfill
\eject
\maketitle

\section{Introduction}
\large
\par    The theory of strong interactions  
states about the prediction of quark-hadron phase transition 
under the condition of 
extreme high nuclear density and very high temperature. 
In the transition phenomena, matter consists of free quarks and gluons called 
quark-gluon plasma (QGP) turns into
 a matter of confined phase of bound quarks of hadrons~\cite{kar,at,sat,bus}. 
The system is believed to exist 
for a short time and the study of this short span
makes us complicated in search of the transition phenomena.
It is believed that the beginning of 
early universe expansion, which is described by Big Bang theory, 
was very hot and subsequently it becomes cool down with the expansion
of the universe. So phase transition
is in due progress of universe's subsequent expansion.
One way to explain the phase transition is through
a process of studying very high temperature system
obtained in the Laboratory. Another
process about the phase transition is through the process of
studying very high nuclear density matter. These very high nuclear density 
matter is normally obtained in
the formation of compact objects, neutron star and 
boson stars~\cite{bub,ru,bh,bh1,dav}. These objects
are believed to be found after the death of giant stars 
in fiery explosion called supernova explosion.
For the investigation and search about the 
nature of the universe, several experimental facilities are 
set up around the globe like
relativistic heavy-ion collision (RHIC) at BNL and large hadron collider
(LHC) at CERN. These two experiments have examined about the creation
and formation of our universe by colliding head on head with very 
energetic ion-beams. These experiments have also claimed for the creation
of mini universe called quark-gluon plasma (QGP)~\cite{c,esh,cai}. 
So these two experiments
give the information about matter at very high temperature. On other hand 
there are some 
experimental facilities like FAIR at Darmstadt and NICA at Dubna, where
the study have focused on dense baryonic matter and the
baryonic matter at Nuclotron (BM@N) experiments which 
extracted ion beams
from modernized Nuclotron, will provide the future information about 
the formation of QGP under the influence of compressed 
dense nuclear matter~\cite{jpc,ba,arsene,back,adams,adcox}.
All the facilities available so far are trying to
detect the existence of the critical point in the phase structure,
the early universe phase transition, formation of QGP 
and chromodynamics (QCD) 
phase structure at very high nuclear density.  
So, the investigation on quark-gluon plasma~(QGP) through the Ultra 
Relativistic Heavy-Ion Collisions has become an exciting field in the current
scenario of heavy ion collider physics.  In this review
article, we focus 
on the calculation of bulk thermodynamic properties of these matter at very
high temperature in continuation of our earlier works of zero loop and one 
loop correction incorporating the two loop correction~\cite{r1,s4}.  
It has been reported more stable droplets of QGP and the corresponding
parametrization value used in the two loop has been largely affected in
the droplet size formation. So the droplet evolution
under two loop correction are more likely to predict the changes in the
stability of droplet formation. 
\section{Potentials of loop correction}
The interacting potential among the quarks and anti-quarks 
through zero-loop is defined in 
the following as

\begin{eqnarray}\label{3.18}
V_{\mbox{zero}}(p) &=& \frac{8 \pi}{p} \gamma~\alpha_{s}(p) T^{2} - \frac{m_{0}^{2}}{2 p}.
\end{eqnarray}
 
Then the potential which is obtained when one loop correction is introduced
in the system.
  
\begin{eqnarray}\label{3.18}
V_{\mbox{one}}(p) &=& V_{zero} [1+\frac{\alpha_{s}(p)a_{1}}{4\pi} ].
\end{eqnarray}

in which $\gamma$ is the parametrization value which is taken in terms of
quark and gluon parametrization factors. The value is different depending on 
zero, one and two loop
corrections~\cite{s1,s2}. In zero-loop 
case the value of quark
and gluon parametrization is $ \gamma_{q}=1/6 $ and $\gamma_{g}=~ (6 - 8)~ \gamma_{q}$
for stable droplet formation whereas it is taken as
$ \gamma_{q}=1/8 $ and $\gamma_{g}=~ (8 - 10)~ \gamma_{q}$ for one loop
correction. So we extend to look for the potential for the case of two loop.
\begin{eqnarray}\label{3.18}
V_{\mbox{two}}(p) &=& V_{\mbox{zero}}[1+\frac{\alpha_{s}(p)a_{1}}{4\pi}+ \frac{\alpha_{s}^2(p)a_{2}}{16 \pi^2}],
\end{eqnarray}

 The value of parametrization for the case of two loop correction to find 
the stable droplet formation
is $ \gamma_{q}=1/14 $ and $\gamma_{g}=~ (48 - 52)~ \gamma_{q}$.
All these factors determine the stability of droplet as well as dynamics of QGP flow similar to the standard Reynold's number of liquid flow and these
parameters support
in forwarding the transformation 
to bound state of matter called hadrons. 
In addition to these parameters we need to define another parameter known as
correction loop parameter $a_{1}$ for one loop in the potential equation, 
which is obtained through the interacting potential among the particles. 
So the coefficient $a_{1}$ is one loop
correction in their interactions and it
is given as~\cite{brambilla,melnikov,hoang}:
\begin{equation}
a_{1}= 2.5833-0.2778~ n_{l},~ 
\end{equation}
where $n_{l}$ is the number of light quark
elements~\cite{fischler,billoire,smirnov,smir}. The value is defined
for one loop correction and we extend our 
calculation into further interactions up to the factor of two loop
correction. In the case of two loop correction the coefficient is obtained as:

\begin{equation}
 a_{2}=28.5468 -4.1471 n_{l}+ 0.0772 n_{l}^2
\end{equation}
 Similarly we have the different mass factors depending on the loop.
The mass of 
the corresponding one loop correction is defined as:
\begin{equation} 
m_{one}^2(T)=2 \gamma^2 g^2(p) T^2[1+g^2(p) a_{1}].
\end{equation} 
whereas the mass of two loop correction
it is: 
\begin{equation} 
m_{two}^2(T)=m_{one}^2(T)~+~2\gamma^2 g^{6}(p) T^2 a_{2}.
\end{equation} 
These thermal masses obtained after incorporation of
one and two loop corrections play pivotal role to find the
interacting potential. 
Really the interaction potentials of loop are probably created due to the 
thermal effective mass of quarks and anti-quarks. Now to obtain
our aim we further look the grand canonical ensemble through 
these loop corrections.

\section{Grand canonical ensemble and free energy} 
We evaluate the grand canonical ensemble through the
loop correction in the interaction potential of quarks, anti-quarks 
by the exchange 
of the coloured particle called gluon. So the evolution obtained
through the ensemble is 
defined in the following through density of state of the
system. The density of state is defined in such way that mean field potential
through the loop correction is incorporated.
The free energies
of quarks, gluons and hadrons can be obtained through the
thermodynamic canonical ensemble of the system.
The general partition function of the system defined by many authors is
given~\cite{satz}
\begin{equation}
Z(T,\mu,V)=Tr\{exp[-\beta(\hat{H}-\mu \hat{N})]\}
\end{equation}

where,$~\mu$ is chemical potential of the system, $\hat{N}$ is quark number and
$\beta =\frac{1}{T}$.
Using this partition function, we correlate the free energy
through the density of states incorporating the loop correction factor, 
and it is
calculated by~\cite{neergaad,ss,ss1}:
\begin{equation}
F_{i}=T\ln Z(T,\mu, V)
\end{equation}
\begin{equation}\label{3.20}
F_i = \eta T g_i \int \rho_{i} (p) p^{2}\ln [1 +\eta e^{-(\sqrt{m_{i}^2 + p^2}-\mu) /T}] dp~,
\end{equation}
where $\rho_{i}$ is the corresponding density of states of loop, 
say~$i=$(zero-loop,~oneloop~and~twoloop),~$\eta=-1$ for the bosonic 
particle and $\eta=+$ for fermionic particles. In the formalism of these 
free energies we use the density of states which
is derived through Thomas and Fermi model in
which the corresponding unloop, one and two loop correction are 
incorporated in the interacting mean field potential.
$g_{i}$ is degree of freedom for quarks and hadronic particles. The value of 
this degree of freedom
is given as $6$ for quarks and $8$ for gluons. It is not a number in the
case of hadronic particles which is defined as
$g_{i}=d v/(2\pi^2)$ where $d$ is number factor like as degree of freedom
of quark and gluon depending on the particular hadronic particle and 
$v=\frac{4}{3}\pi r^3$ represents as the volume of the hadron droplet.
However, the density of states~$\rho_{i}$ for hadronic particles
is unity with its momentum
factor $p^{2}$ and it is defined for the case of quarks and gluon in a way 
that mean field potential
through the loop correction is incorporated 
with unit momentum factor and corresponding density of states~\cite{ss3}. 
The density of states for quarks and gluon
for the corresponding loops are defined below and when the loop is 
incorporated, the corresponding density of state is applied in 
the calculation of their free energies using the corresponding 
mean field potential.
\begin{equation}
\rho_{i} (p) = \frac{v}{3 \pi^{2}}\frac{dV^{3}_{conf}(p)}{dp}~,
\end{equation}
or,
\begin{equation}\label{3.13}
\rho_{zero}(p) =\frac{\nu}{\pi^2}\frac{\gamma^{9}T^6}{8} g^{6}(p) A, 
\end{equation}

where,
\begin{eqnarray}
A&=&\frac{1}{p^2}[\frac{1}{p^2}+\frac{2}{(p^2+\Lambda^2)\ln(1+\frac{p^2}{\Lambda^2})}]. 
\end{eqnarray}
whereas in the case of one loop correction then the density of state 
is obtained as:
\begin{equation}\label{3.13}
\rho_{one}(p) =\frac{\nu}{\pi^2}\frac{\gamma^{9}T^6}{8} g^{6}(p) B, 
\end{equation}
where
\begin{eqnarray}
B&=&\lbrace 1+\frac{\alpha_{s}(p)a_{1}}{\pi}\rbrace^{2}[ \frac{(1+\alpha_{s}(p)a_{1}/\pi)}{p^{4}} \nonumber \\
&+&\frac{2 (1+2\alpha_{s}(p)a_{1}/\pi)}{p^{2}(p^2+\Lambda^2)\ln(1+\frac{p^2}{\Lambda^2})}] 
\end{eqnarray}
Similarly, we obtain the density of state for two loop system as follows:
\begin{equation}\label{3.13}
\rho_{two}(p) =\frac{\nu}{\pi^2}\frac{\gamma^{9}T^6}{8} g^{6}(p) C, 
\end{equation}

where
\begin{eqnarray}
C&=&[ 1+\frac{\alpha_{s}(p)a_{1}}{\pi}+\frac{\alpha_{s}^2(p)a_{2}}{\pi^2}]^{2}
\nonumber \\
&\times&[ \frac{(1+\alpha_{s}(p)a_{1}/\pi 
+\alpha_{s}(p)^2a_{2}/\pi^2)}{p^{4}} \nonumber \\
&+&\frac{2 (1+2\alpha_{s}(p)a_{1}/\pi+3 \alpha_{s}(p)^2a_{2}/\pi^2)}{p^{2}(p^2+\Lambda^2)\ln(1+\frac{p^2}{\Lambda^2})}] 
\end{eqnarray}
%
%\begin{equation}
%\rho_{i} (p) = \frac{v}{3 \pi^{2}}\frac{dV^{3}_{conf}(p)}{dp}~,
%\end{equation}
%or,
%\begin{equation}\label{3.13}
%\rho_{two}(p) =\frac{\nu}{\pi^2}\frac{\gamma^{9}T^6}{8} g^{6}(p) A, 
%\end{equation}

%where
%\begin{eqnarray}
%A&=&[ 1+\frac{\alpha_{s}(p)a_{1}}{\pi}+\frac{\alpha_{s}^2(p)a_{2}}{\pi^2}]^{2}
%\nonumber \\
%&\times&[ \frac{(1+\alpha_{s}(p)a_{1}/\pi 
%+\alpha_{s}(p)^2a_{2}/\pi^2)}{p^{4}} \nonumber \\
%&+&\frac{2 (1+2\alpha_{s}(p)a_{1}/\pi+3 \alpha_{s}(p)^2a_{2}/\pi^2)}{p^{2}(p^2+\Lambda^2)\ln(1+\frac{p^2}{\Lambda^2})}] 
%\end{eqnarray}
%and~ $v$ is the volume occupied by the QGP and
%$~g^{2}(p)=4 \pi \alpha_{s}(p)$~ and if the two loop correction is removed,
%then the value of $B$ is reduced to one loop correction and become $A$,
%which is given as:
%\begin{equation}\label{3.13}
%\rho_{oneloop}(p) =\frac{\nu}{\pi^2}\frac{\gamma^{9}T^6}{8} g^{6}(p) B, 
%\end{equation}
%where
%\begin{eqnarray}
%B&=&\lbrace 1+\frac{\alpha_{s}(p)a_{1}}{\pi}\rbrace^{2}[ \frac{(1+\alpha_{s}(p)a_{1}/\pi)}{p^{4}} \nonumber \\
%&+&\frac{2 (1+2\alpha_{s}(p)a_{1}/\pi)}{p^{2}(p^2+\Lambda^2)\ln(1+\frac{p^2}{\Lambda^2})}] 
%\end{eqnarray}
%Again if we put $a_{1}=0$ then the factor "A" be reduced to the 
%factor "C" which is equal and defined as

%\begin{equation}\label{3.13}
%\rho_{unloop}(p) =\frac{\nu}{\pi^2}\frac{\gamma^{9}T^6}{8} g^{6}(p) C, 
%\end{equation}
%where,
%\begin{eqnarray}
%C&=&\frac{1}{p^2}[\frac{1}{p^2}+\frac{2}{(p^2+\Lambda^2)\ln(1+\frac{p^2}{\Lambda^2})}]. 
%\end{eqnarray}

It implies that due to the correction factor, the density of states
is perturbed by small factor which can be seen in the figure of potential
vs momentum. In the expression, the parameter $~\Lambda $ is considered 
in the scale of QCD as
~$ 0.15~$GeV. So we can set up the free energy of the system by
finding the energies of quarks, anti-quarks, gluons, all the light
and medium light hadrons.  
The integral is evaluated from the least value of momentum approximately 
tending to zero. Taking and considering  all the massive 
hadrons now the total free energy is calculated by adding the inter-facial 
energy of the fireball~\cite{ss2,ss4}. 

\begin{equation}
 F_{total}=\sum_{i} F_{i}~+~\frac{\gamma T R^{2}}{4}\int p^2 \delta(p-T) dp, 
\end{equation}

~ in the first term of the total free energy the summation in which 
$i$ stands for $u$,~$d$,~$s$~quarks and all the hadronic 
particles with gluon whereas in second term, it is inter-facial energy
which replace the role of bag energy of MIT bag model in which Bag energy was
introduced in the scale of $B^{1/4}=T_{c}$. Taking
the inter-facial energy in place of MIT Bag energy, it can reduce
the drawback produced by Bag energy to the maximum effects in 
comparison to MIT model calculation and this inter-facial energy is 
dependent on temperature and some parametric factor. So in the
inter-facial energy, $R$ is size of QGP droplet with 
the parametrization factor.

%%%%%%%%%%%%%
\begin{figure}[htb]
\centering
%\sidecaption
\includegraphics[width=7cm,clip]{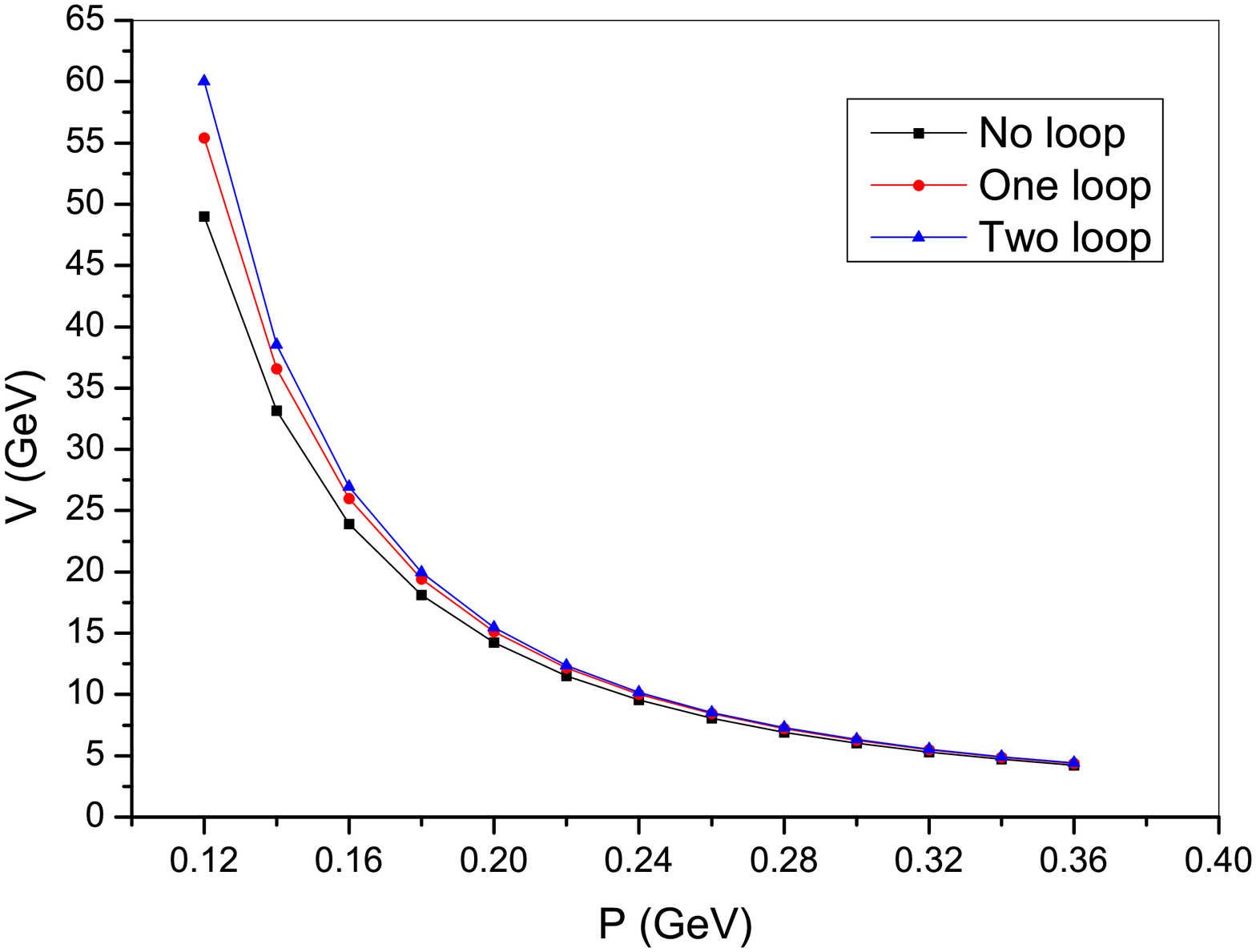}
%\vspace*{5cm}       % Give the correct figure height in cm
\caption{  Potential ~ vs.~Momentum~with and without loop.}
\label{fig-2}       % Give a unique label
\end{figure}
%%%%%%%%%%%%
\begin{figure}[htb]
\centering
%\sidecaption
\includegraphics[width=7cm,clip]{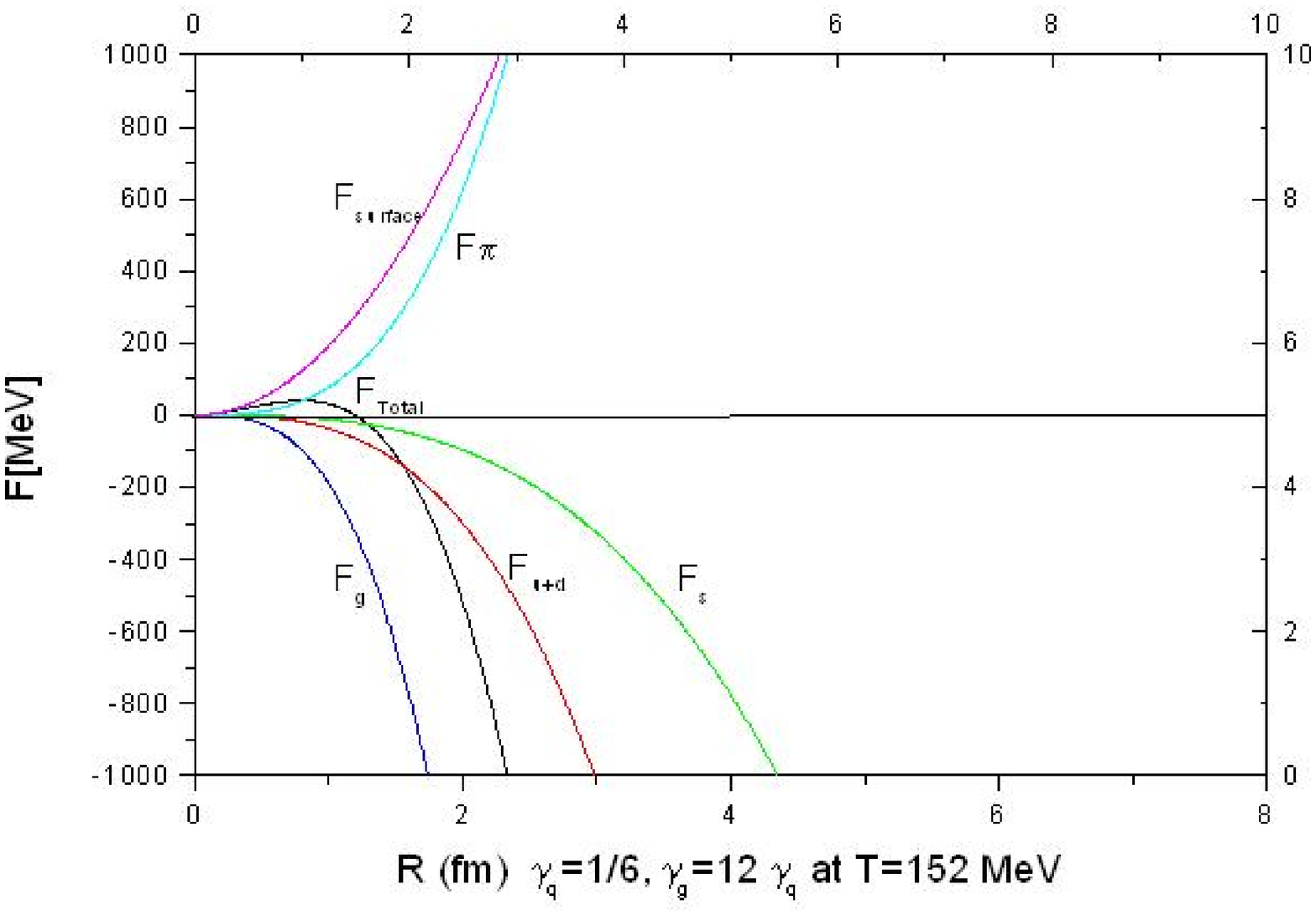}
\caption{Free energy vs.~$R$~
at~$T=152$~MeV for zero-loop for contribution particles.}
\label{fig-4}       % Give a unique label
\end{figure}

%%%%%%%%%%%%%%%%%
\begin{figure}[htb]
\centering
%\sidecaption
\includegraphics[width=7cm,clip]{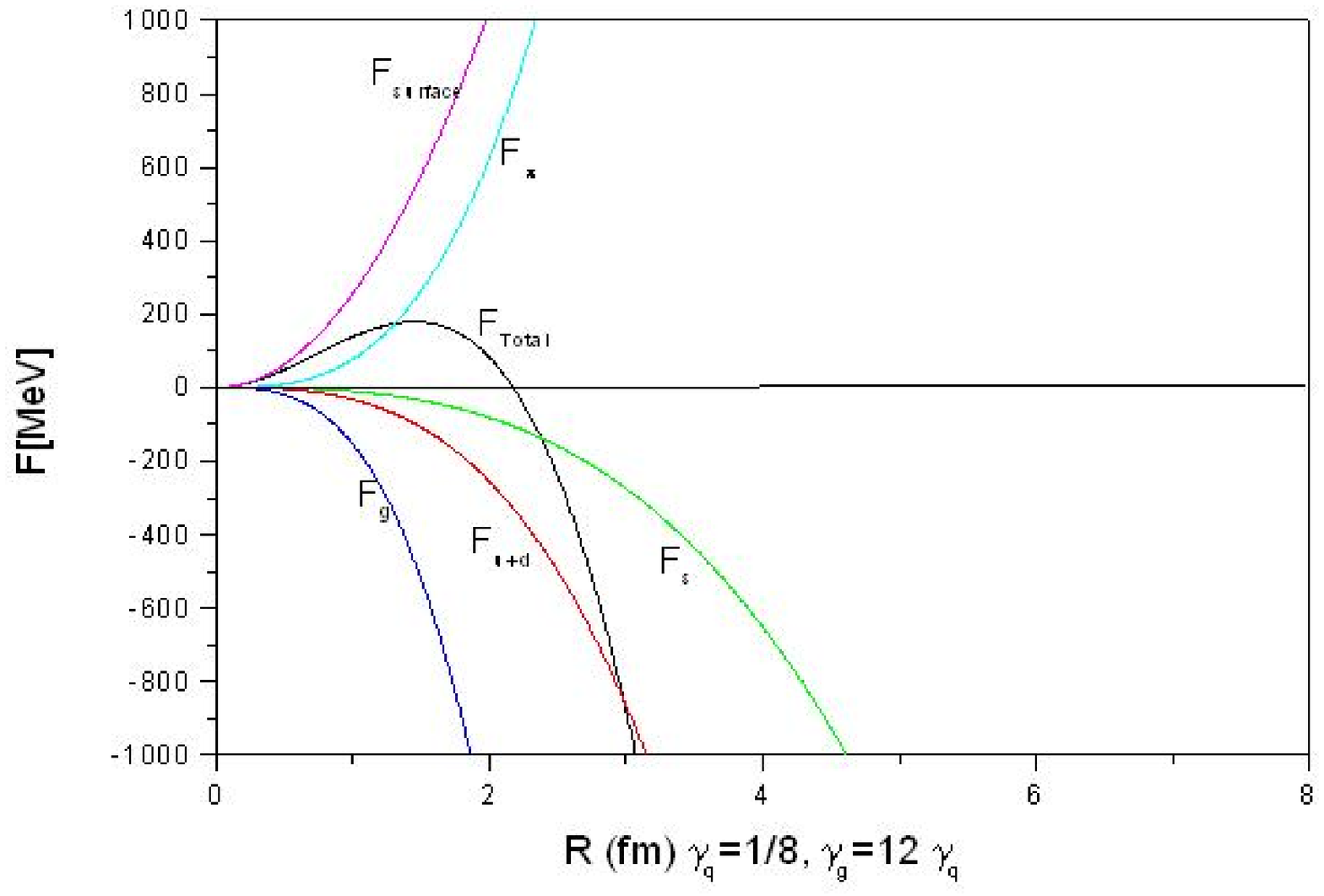}
\caption{Free energy vs.~$R$~
at~$T=152$~MeV for one loop for contribution particles.}
\label{fig-4}       % Give a unique label
\end{figure}
%%%%%%%%%%%
\begin{figure}[htb]
\centering
%\sidecaption
\includegraphics[width=7cm,clip]{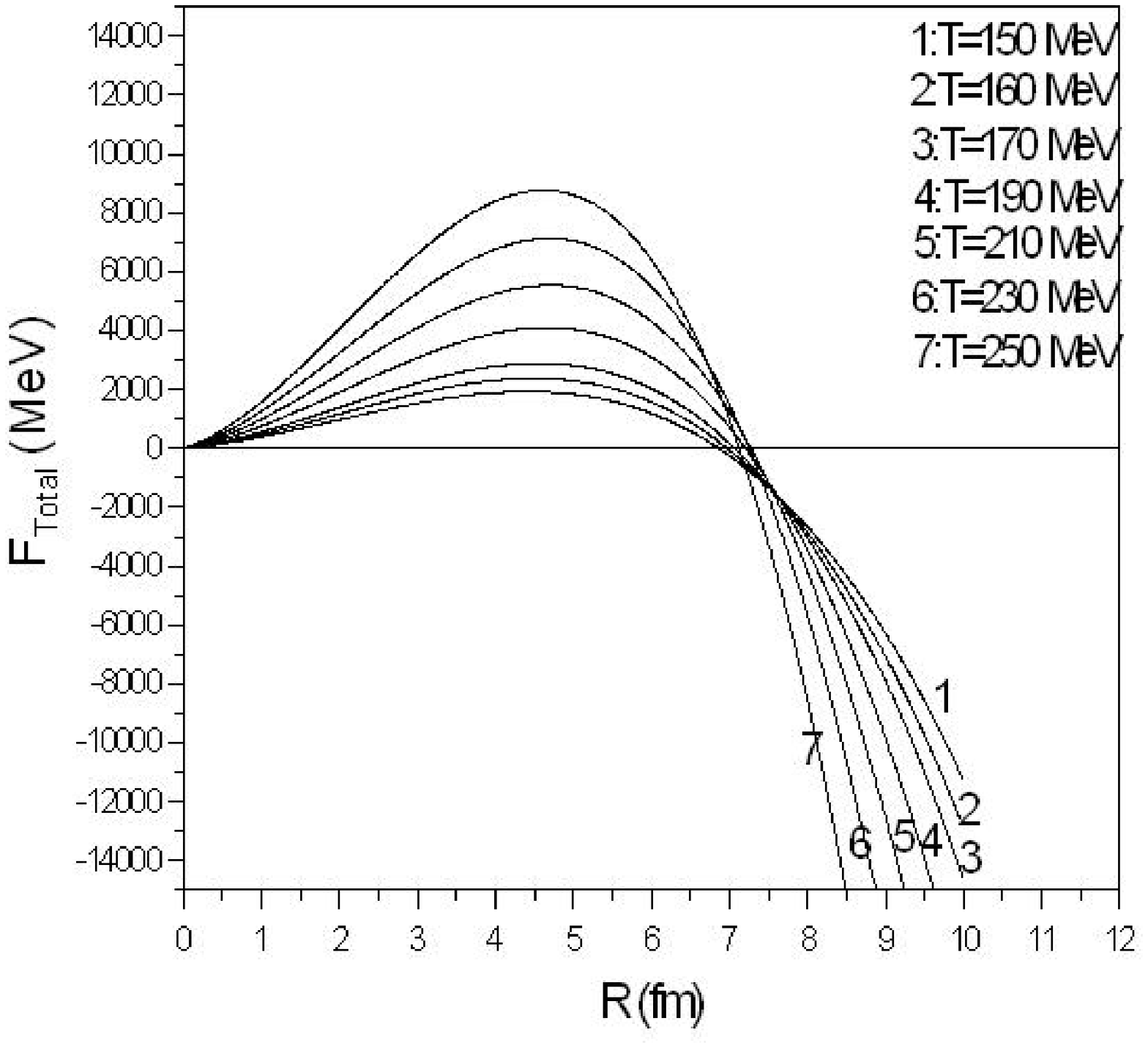}
%\vspace*{5cm}       % Give the correct figure height in cm
\caption{  Free Energy~ vs.~$R$ (fm)~at~$\gamma_{q}=1/6~$, $\gamma_{g}=6\gamma_{q}$ for zero-loop correction.}
\label{fig-2}       % Give a unique label
\end{figure}
%%%%%%
\begin{figure}[htb]
\centering
%\sidecaption
\includegraphics[width=7cm,clip]{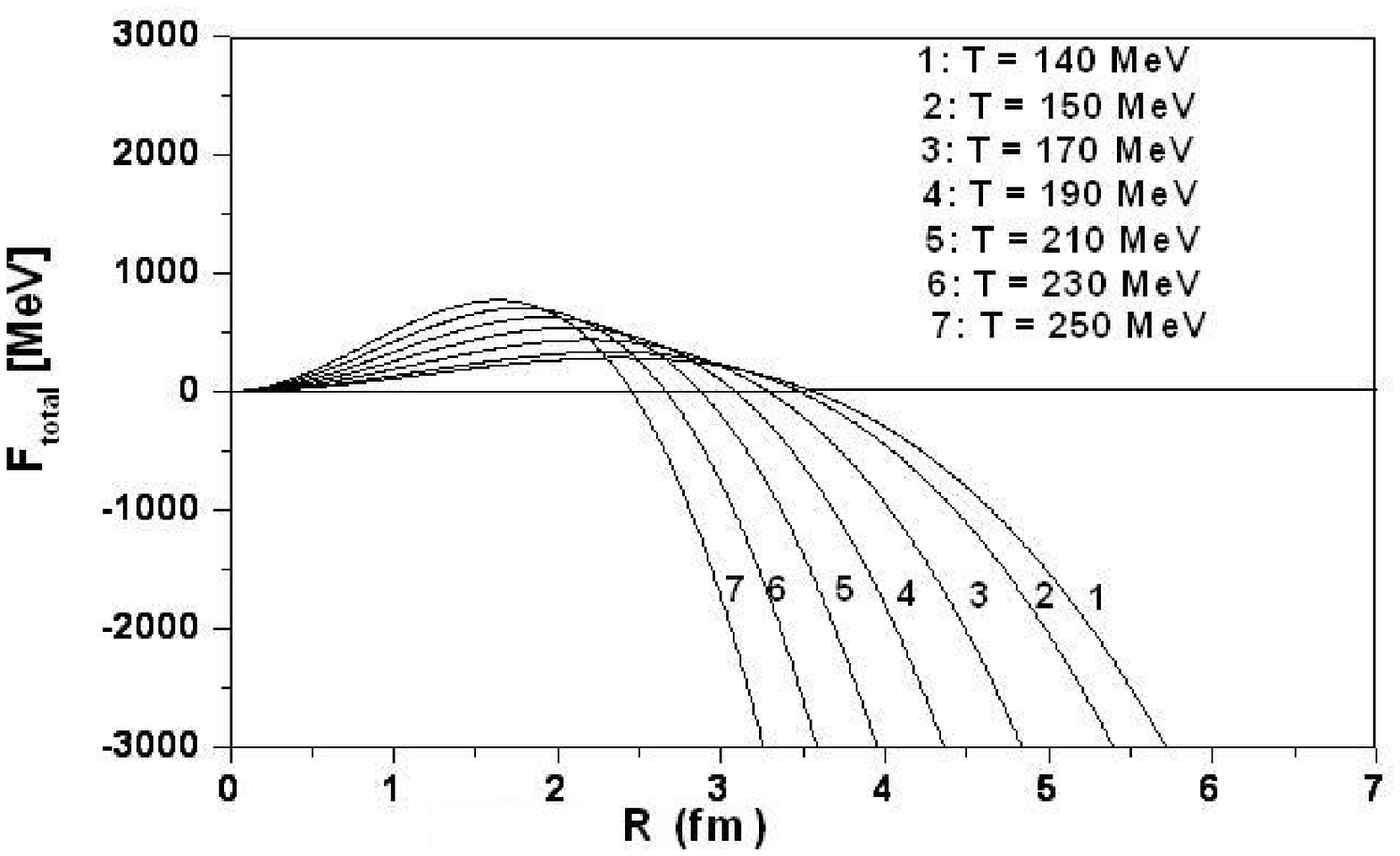}
%\vspace*{5cm}       % Give the correct figure height in cm
\caption{  Free Energy~ vs.~$R$ (fm)~at~$\gamma_{q}=1/6~$, $\gamma_{g}=8\gamma_{q}$ for zero-loop correction.}
\label{fig-2}       % Give a unique label
\end{figure}
%%%%%%%%
\begin{figure}[htb]
\centering
%\sidecaption
\includegraphics[width=7cm,clip]{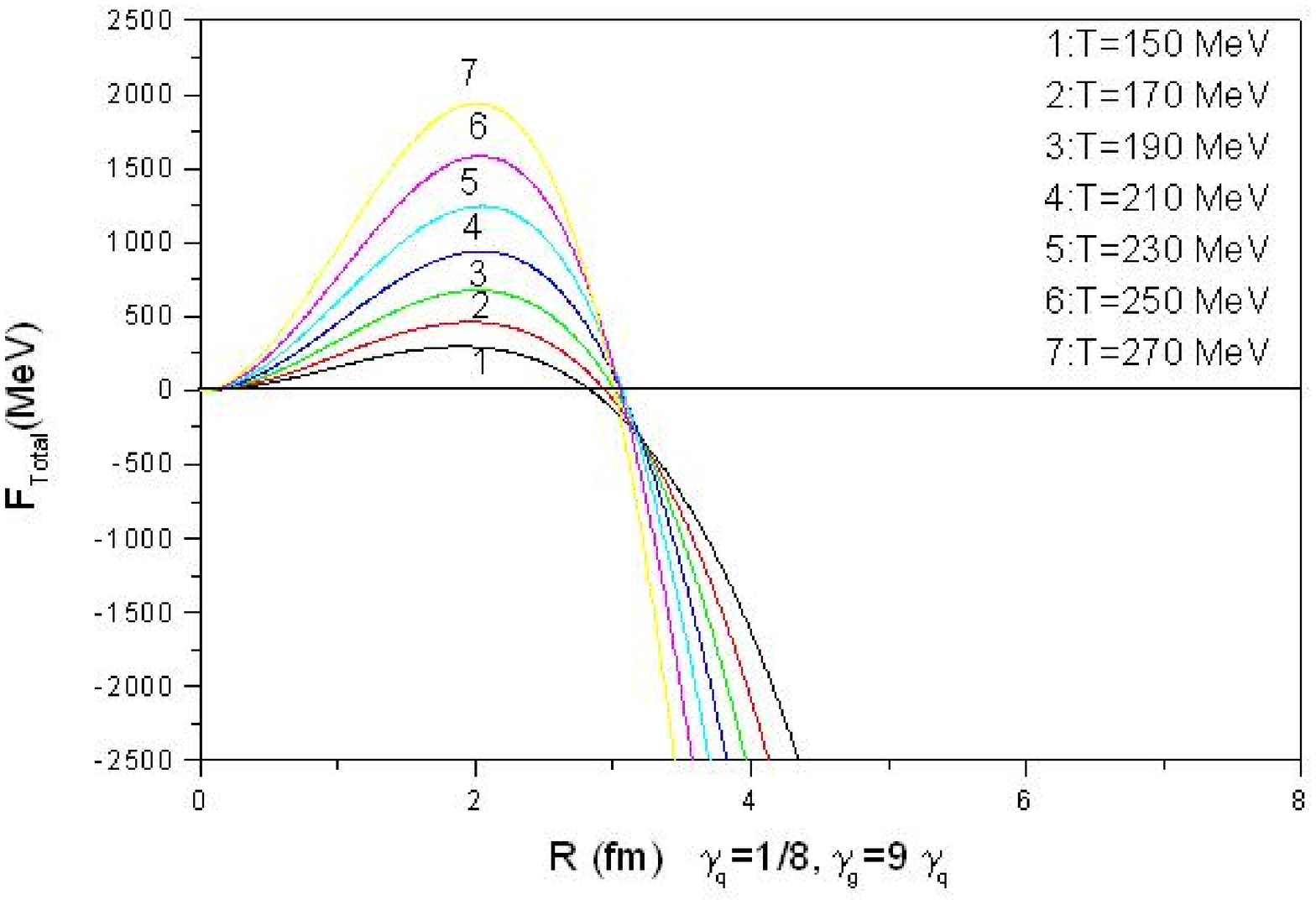}
%\vspace*{5cm}       % Give the correct figure height in cm
\caption{  Free Energy~ vs.~R (fm)~at~$\gamma_{q}=1/8~$, $\gamma_{g}=9\gamma_{q}$for one loop correction}
\label{fig-2}       % Give a unique label
\end{figure}
%%%%%%%%%%%%%%
\begin{figure}[htb]
\centering
%\sidecaption
\includegraphics[width=7cm,clip]{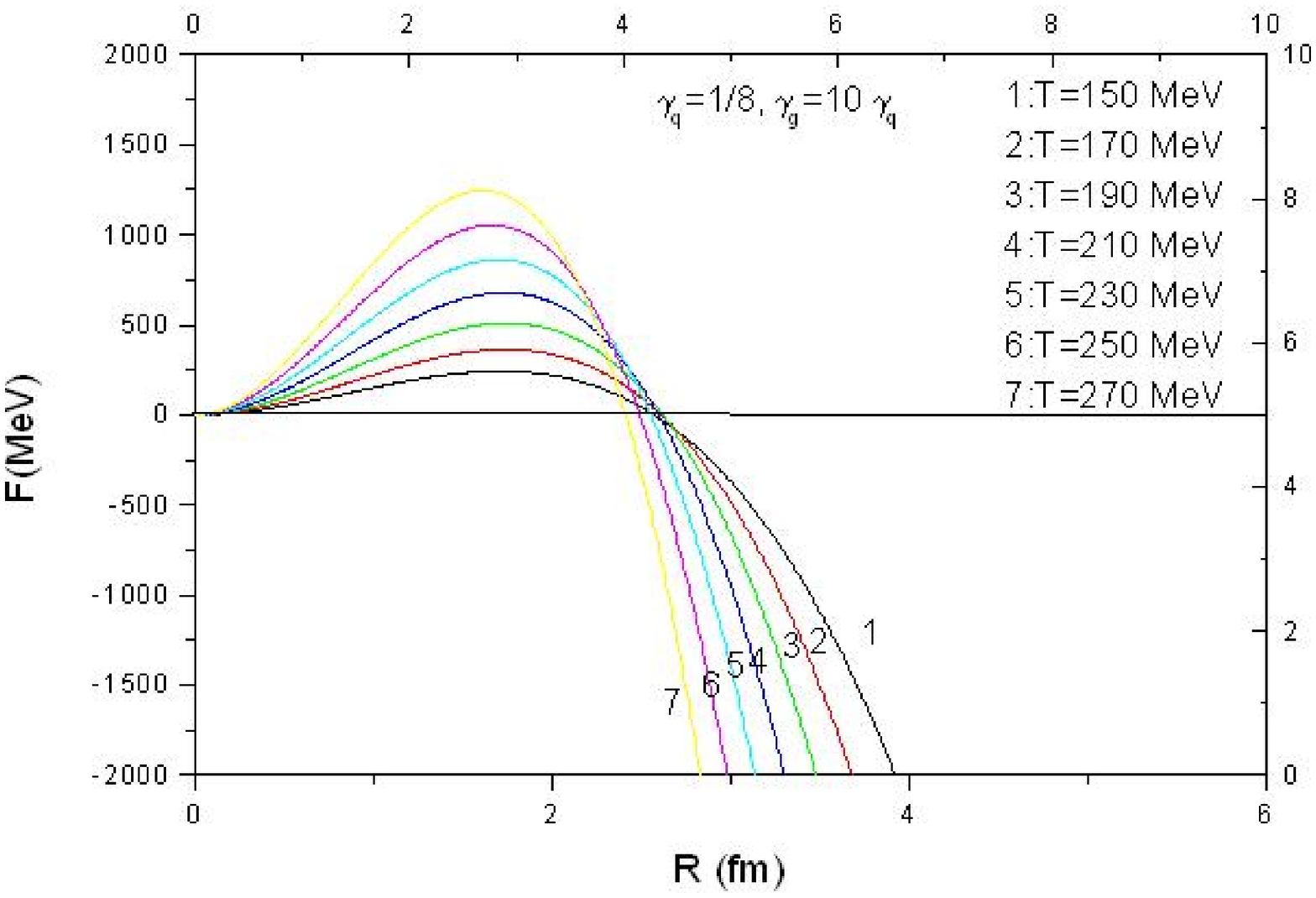}
%\vspace*{5cm}       % Give the correct figure height in cm
\caption{  Free Energy~ vs.~R (fm)~at~$\gamma_{q}=1/8~$, $\gamma_{g}=10\gamma_{q}$for one loop correction}
\label{fig-2}       % Give a unique label
\end{figure}
%%%%
\begin{figure}[htb]
\centering
%\sidecaption
\includegraphics[width=7cm,clip]{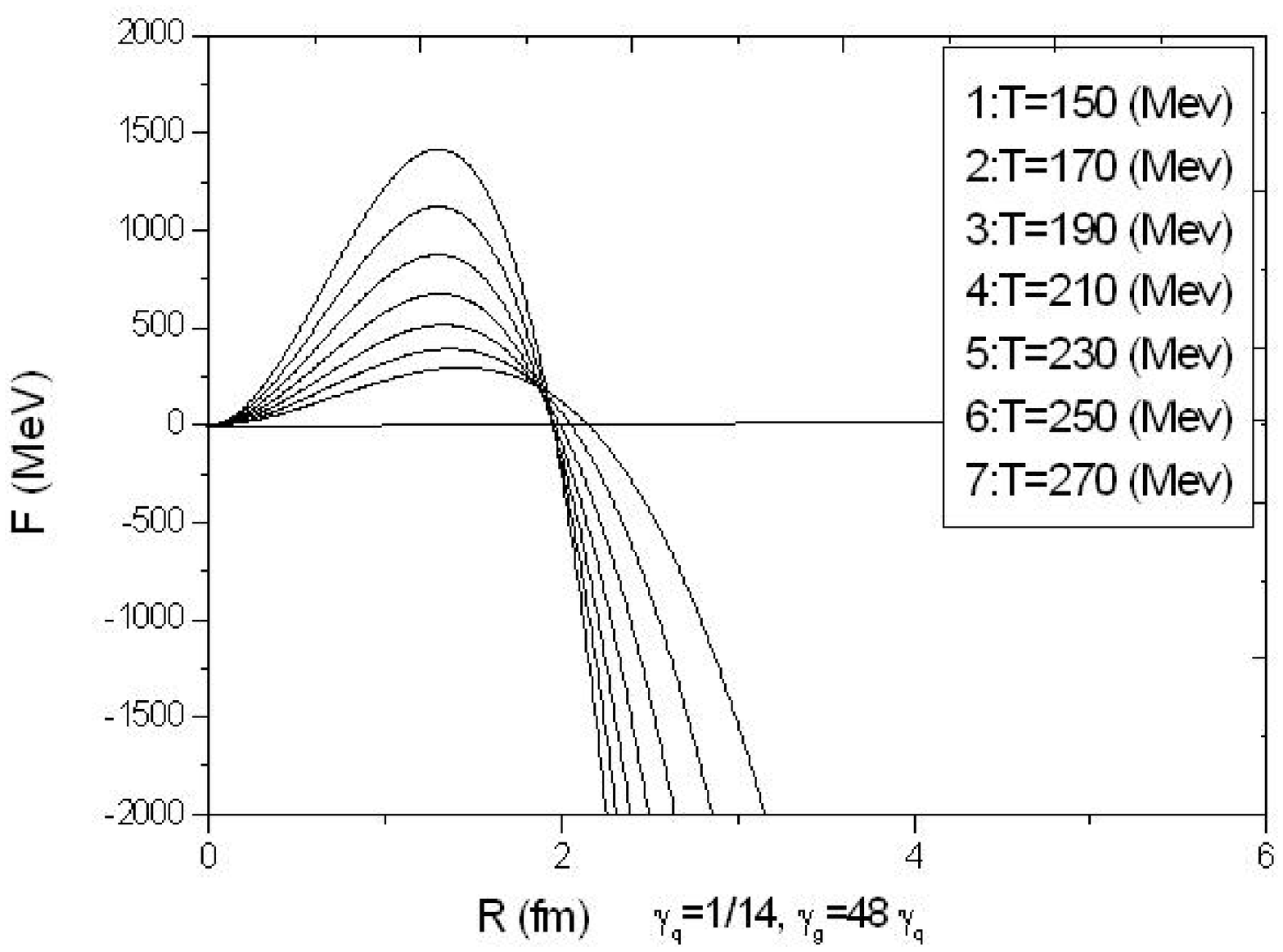}
%\vspace*{5cm}       % Give the correct figure height in cm
\caption{ Free Energy~ vs.~R (fm)~at~$\gamma_{q}=1/14~$, $\gamma_{g}=48\gamma_{q}$ for two loop correction.}
\label{fig-2}       % Give a unique label
\end{figure}
%%%%%%%%%%%%
\begin{figure}[htb]
\centering
\includegraphics[width=7cm,clip]{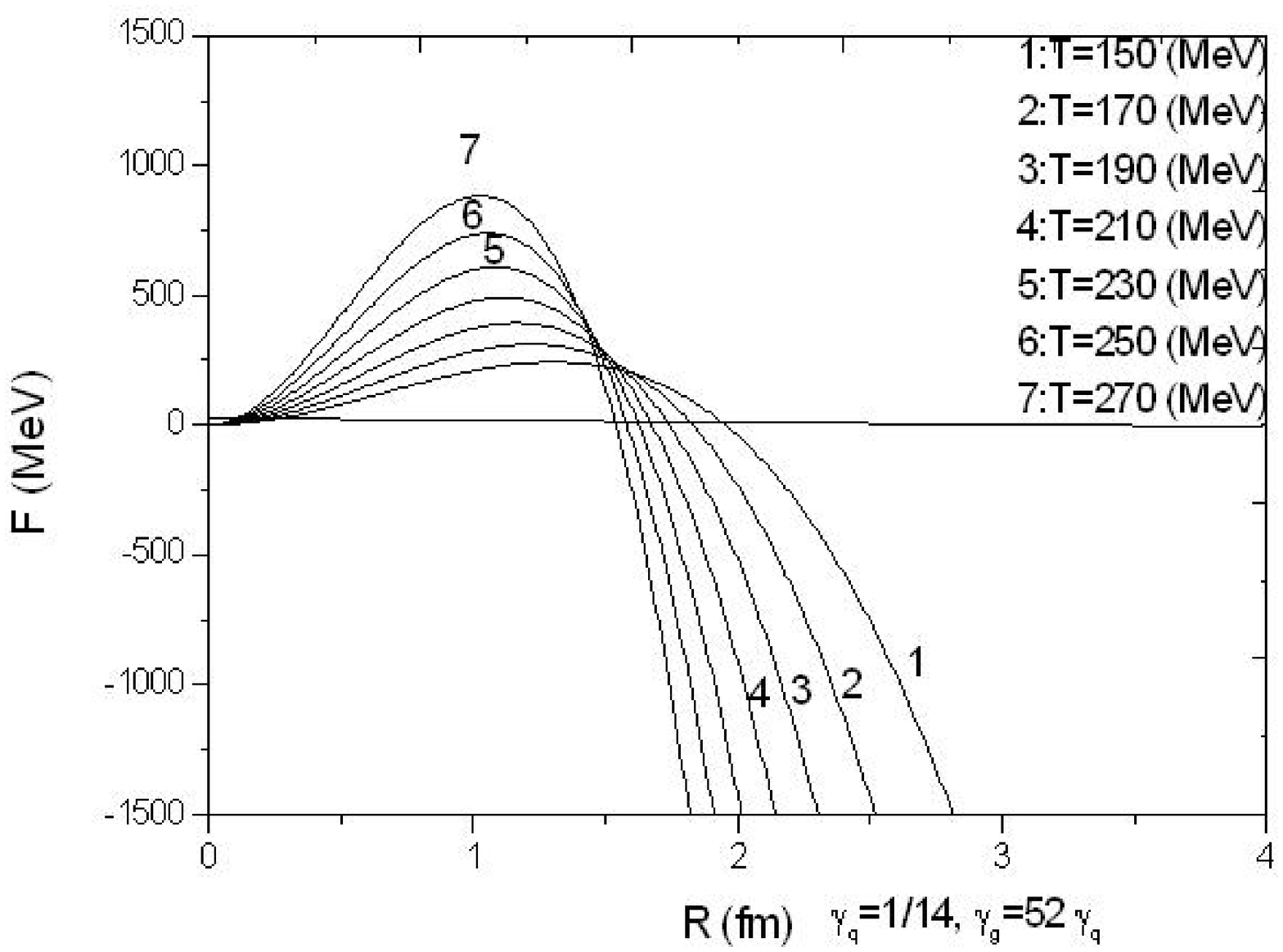}
%\vspace*{5cm}       % Give the correct figure height in cm
\caption{ Free Energy~vs.~R (fm)~at~$\gamma_{q}=1/14~$, $\gamma_{g}=52\gamma_{q}$ for two loop correction.}
\label{fig-2}       % Give a unique label
\end{figure}
%%%%%%%%%%
\section{Results:} The analytical calculations of free energy
of QGP-hadron fireball evolution without and with one and two loop
correction factors in the interacting mean-field potential are performed
by computing the variation of the interacting potential and momentum. The 
potential function is slightly perturbed from the unloop factor. 
These characteristic feature is shown in Fig.$1$. By the loop
correction it is slightly increased at the lower momentum region and with
increasing momentum, the perturbative contribution is negligible 
indicating that the perturbations of the one and two loop correction 
are very small in the 
high momentum transfer. Then we look forward the free energy change with
droplet size of the different contributed particles for the unloop potential 
at a particular temperature say $T=152$~MeV as ad-hoc assumption. 
At this particular temperature $T=152~$MeV, we again
look the change in behaviour of free energy of the constituted particles
for one loop potential as considered in case of unloop
potential and we obtain the similar behaviour with a
slight difference in free energy amplitude. It implies that there is similar
phenomenon of zero-loop and loop in the free energy graph with different 
amplitudes at any particular temperature. 
\par Now in Fig.$4$, we show free energy
evolution for zero-loop potential for different temperatures 
at parametrization $\gamma_{q}=1/6$ and $\gamma_{g}=6 \gamma_{q}$.
This particular parametrization $\gamma_{q}=1/6$                     
and $\gamma_{g}=6 \gamma_{q}$ which we chosed is due to the
finding of the stable droplet. The value is
obtained after ad-hoc search of stable droplet. So 
there is only one stable
droplet formation for unloop potential. If we further look stable droplet
by increasing the parameter
beyond at $\gamma_{q}=1/6$                     
and $\gamma_{g}\ge 6 \gamma_{q}$, there is no more stable droplet at any other
parametrization value which is represented by Fig.$5$.
It means the stable droplet formation is
obtained specially at parametrization value say $\gamma_{q}=1/6$~
and~$\gamma_{g}=6 \gamma_{q}$ and other
droplets are not exactly stable
in unloop potential even though droplets be formed. Beyond $\gamma_{g}>6 \gamma_{q}$ there is droplet but
no stability. It denotes that the parameter value behaves like
a representation to control in the behaviour of fluid dynamics 
in terms of its stable 
droplet formation. 
However when the value of $\gamma_{q}$ is not $1/6$ and $\gamma_{g}$ is not
equal to six times of $\gamma_{q}$ then we can have unstable QGP formation and
QGP fluid may be any sorts of unpredictable fluid dynamics.
In Fig.$6$ and $7$, there are free energy                      
representations of one loop showing smaller droplet size with different
temperatures.
It indicates that we have found similar stable droplets 
at the parametrization values
$\gamma_{q}=1/8$ and $8 \gamma_{q}\le \gamma_{g} \le 10 \gamma_{q}$ in the case of additional oneloop potential.
\par In Fig.$6$ it shows changing slightly in the
stable droplet size from droplet of unloop potential. Similarly
in Fig.$7$ stable droplet is available to obtain by changing
the parametrization value to $\gamma_{g}\le 10 \gamma_{q}$.
From these two presentations,
there are almost around two stable droplets formations and 
these stability's are only found
in the range of parametrisation $8 \gamma_{q}\le \gamma_{g}\le 10\gamma_{q}$.
In a similar way, we look phenomenologically about the evolution of free energy
by adding twoloop correction in the potential. The behaviour of droplets 
are shown in Fig.$8$ and Fig.$9$. The
behaviour of evolution of free energy with the increase of QGP drop size is
much smaller in comparison to earlier droplets obtained by unloop
and one loop correction in the potential.
The addition of such two loop
correction shows the formation of stable droplet modified a lot and 
there is change in the parameter factor also. The stable droplet is 
still observed for different parameter values. The values are found 
to be $\gamma_{q}=1/14$ and $48 \gamma_{q} \le \gamma_{g}\le 52\gamma_{q}$. 
Only at these values we can 
observe the stable QGP droplet for two loop correction. Beyond these 
parametrization, QGP fluid remains like unstable behaviour of zero-loop type 
with having different in-oriental 
fluid dynamics and represent unstable droplets. These characteristic features
are all described in the corresponding figures $8$ and $9$.
With the incorporation of two loop, the droplet size is much smaller 
indicating highly stable. 
The smaller the size, surface tension of drop is large and the
liquid drops are tightly bounded so that the droplet is more stable. 
So parametrization factors for zero-loop, one loop and two loop play a very
much important role in finding the stable droplet in the evolution of QGP
droplet formation and it is really significant parameter for formation
of QGP droplets.

\section{conclusion:}
We can conclude from these results that due to the presence of
loop corrections in the mean field potential, 
the stability is found to be much better in the case of 
two loop correction by the smaller size of droplet which is not 
comparable in terms of one loop correction and without loop 
correction where sizes of droplets are found to be bigger. So loop 
corrections in the potential can be studied as phenomenological model
for describing the droplet formation of QGP taken as a 
dynamical parameter.
\subsection{\bf Acknowledgments:}

We are very much thankful to our retired Prof. R Ramanathan for his 
untiring work in support of preparing the manuscript with his critical 
reading and many discussion about the possible outcomes of the manuscript.

\end{document}